%% file: main.tex
\documentclass[fleqn,10pt]{wlscirep}
\usepackage[utf8]{inputenc}
\usepackage[T1]{fontenc}
\usepackage{lineno}

\usepackage{siunitx} 
\DeclareSIUnit{\volpercent}{vol\%}
\input{Images/flowchart_tikz_headers}
\usepackage{standalone}
\usepackage{epsfig}
\usepackage{cleveref}

\title{The BELSAR dataset: Mono- and bistatic full-pol L-band SAR for agriculture and hydrology} 

\author[1,*,$\dag$]{Jean~Bouchat}
\author[2,$\dag$]{Emma~Tronquo} 
\author[3]{Anne~Orban}
\author[4]{Karlus~A.C.~de~Macedo}
\author[2]{Niko~E.~C.~Verhoest}
\author[1]{Pierre~Defourny}

\affil[1]{Earth and Life Insitute, Université catholique de Louvain, Louvain-la-Neuve, 1348, Belgium}
\affil[2]{Hydro-Climate Extremes Lab, Ghent University, Ghent, 9000, Belgium}
\affil[3]{Centre Spatial de Liège, Université de Liège, Angleur, 4031, Belgium}
\affil[4]{MetaSensing B.V., DK Noordwijk, 2201, The Netherlands}

\affil[*]{corresponding author: Jean Bouchat (jean.bouchat@uclouvain.be)}
\affil[$\dag$]{these authors contributed equally to this work}

\begin{abstract}
The BELSAR dataset is a unique collection of high-resolution airborne mono- and bistatic fully-polarimetric synthetic aperture radar (SAR) data in L-band, alongside concurrent measurements of vegetation and soil bio-geophysical variables measured in maize and winter wheat fields during the summer of 2018 in Belgium. This innovative dataset, the collection of which was funded by the European Space Agency (ESA), helps addressing the lack of publicly-accessible experimental datasets combining multistatic SAR and in situ measurements. As such, it offers an opportunity to advance the development of SAR remote sensing science and applications for agricultural monitoring and hydrology. This paper aims to facilitate its adoption and exploration by offering comprehensive documentation and integrating its multiple data sources into a unified, analysis-ready dataset.
\end{abstract}

\begin{document}

\flushbottom
\maketitle

\thispagestyle{empty}


\section*{Background \& Summary}


Agriculture and soil moisture monitoring are essential for sustainable food production, water resource management and mitigating the impact of climate change on crop yield and ecosystem health.
Today, the best-established large-scale operational agricultural monitoring systems are mainly based on optical remote sensing \cite{mcnairn2016review}. However, they are often hampered by the presence of clouds \cite{whitcraft2015cloud}, which obscure the view of the sensors. In contrast, Synthetic Aperture Radars (SAR) are active sensors, largely independent of weather conditions, capable of interacting with and capturing valuable information about both vegetation canopies and underlying soils \cite{el2018penetration,vreugdenhil2018sensitivity}.
The recent European Space Agency (ESA) Sentinel-1 (S1) mission, providing systematic and global coverage of dense SAR time series, paved the way for the development of operational applications like the Copernicus Emergency Services and the open-source Sen4CAP system to implement the European Union's Common Agricultural Policy. The success of this operational mission triggered the intent to launch S1 companion satellite to enhance this all-weather earth-observation capacity in a bistatic mode.

Bistatic SARs are radar systems in which the transmitter and receiver are physically separated, unlike monostatic systems in which they share the same location. Multistatic systems, meanwhile, have several receivers for a common transmitter. The most simple multistatic system comprises of both an active monostatic sensor and a passive bistatic one. These systems allow the acquisition of images using different geometries and configurations, thereby capturing multi-dimensional scattering effects that could not be recorded by monostatic systems alone.
These flexible systems offer the advantage of helping to differentiate between the often intertwined relative contributions of vegetation and soil, thus improving the retrieval performances of soil moisture and crop biophysical variables retrieval \cite{guerriero2013use}.
Despite its considerable potential, bi- or multistatic SAR applications for vegetation and soil monitoring have remained limited. It seems that this scarcity can mostly be attributed to a lack of comprehensive, well-documented experimental datasets combining mono- and bistatic SAR acquisitions and in situ measurements of vegetation and soil biogeophysical variables \cite{pierdicca2021retrieval}. As a result, the potential of bistatic SAR for these applications has mainly been investigated via radiative transfer models \cite{pierdicca2021retrieval,zhang2016bistatic,zeng2016radar,guerriero2013use,della2006advances}.

In this context, the ESA funded the BELSAR-Campaign project, an airborne and in situ measurement campaign that took place during the 2018 growing season in Belgium. The result is the BELSAR dataset \cite{belsar2018}, a collection of data containing high-resolution fully-polarimetric mono- and bistatic synthetic aperture radar (SAR) times series in L-band and concurrent field measurements of vegetation and soil bio-geophysical variables. The SAR data were acquired with an airborne multistatic SAR system operated by MetaSensing BV. The field measurements were collected both during and after the crop growing season in ten maize and ten winter wheat fields simultaneously with the SAR acquisitions.

Several studies exploiting the BELSAR dataset have already been published. 
Bouchat \textit{et al.} \cite{bouchat2022assessing} have used the dataset to assess the potential of simultaneous mono- and bistatic SAR acquisitions for agriculture and soil moisture monitoring applications, as well as the impact of maize row structure on the SAR backscatter.
Tronquo \textit{et al.} \cite{tronquo2022soil} have presented a semi-empirical method based on effective roughness modeling to retrieve soil moisture in bare agricultural fields and have shown an increase in the accuracy of soil moisture estimation by using several polarizations at the same time. 
Finally, Bouchat \textit{et al.} \cite{bouchat2022green} have obtained promising results for green area index (GAI) retrieval in maize fields using the Water Cloud Model \cite{prevot1993estimating} and dual-polarized SAR data in L-band.
Yet, the potential of the BELSAR dataset is still far from having been fully investigated and, given the vast possibilities offered by such an innovative and unique collection of data, other users will want to exploit it in their research in agriculture, hydrology, change detection, or other SAR remote sensing techniques and applications. Therefore, this paper aims to facilitate its uptake through its thorough documentation, as well as through the integration of the different sources of data from BELSAR-Campaign into a single so-called integrated dataset provided with it.

\section*{Methods}


\subsection*{Airborne SAR acquisitions}

\begin{figure}[h]
    \centering
    \includegraphics[width=.65\linewidth]{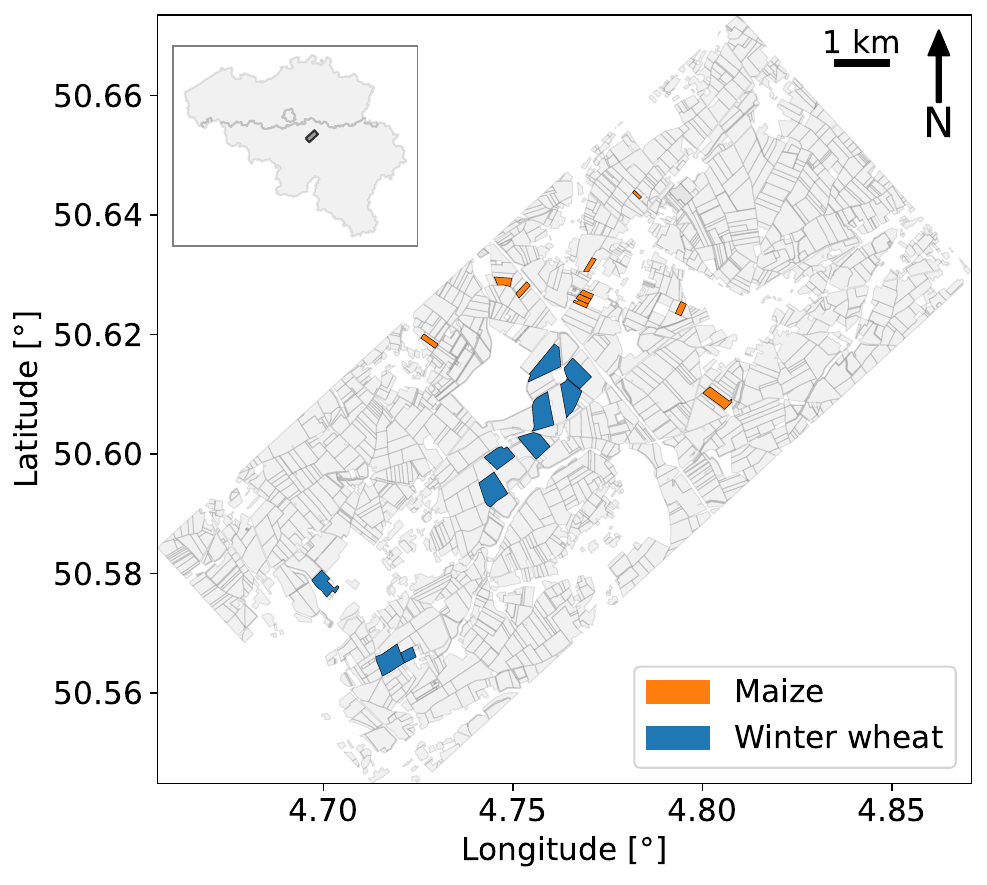}
    \caption{BELSAR area of interest, with the ten winter wheat and ten maize fields in which the in situ measurements were recorded in blue and orange respectively, and all other agricultural fields inventoried in the Land Parcel Information System of Wallonia, Belgium, in grey.}
    \label{fig:BELSAR_AOI}
\end{figure}

The design and implementation of the airborne SAR campaign was carried out by the Centre Spatial de Liège and MetaSensing BV.
Multi-temporal mono- and bistatic fully-polarimetric (HH, HV, VH, and VV) airborne L-band SAR data were acquired with the MetaSAR-L systems over the BELSAR area of interest, in \cref{fig:BELSAR_AOI}, and processed with the MetaSAR-Pro software, MetaSensing BV's proprietary airborne SAR processor.

\begin{figure}[!t]
    \begin{minipage}[b]{.30\linewidth}
        \centering
        \centerline{\epsfig{figure=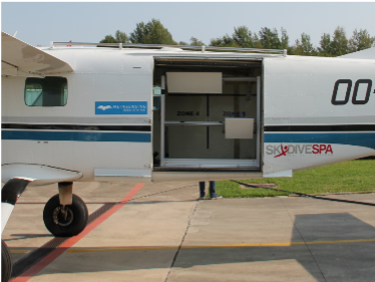, width=\linewidth}}
    \end{minipage}
    \hfill
    \begin{minipage}[b]{.30\linewidth}
        \centering
        \centerline{\epsfig{figure=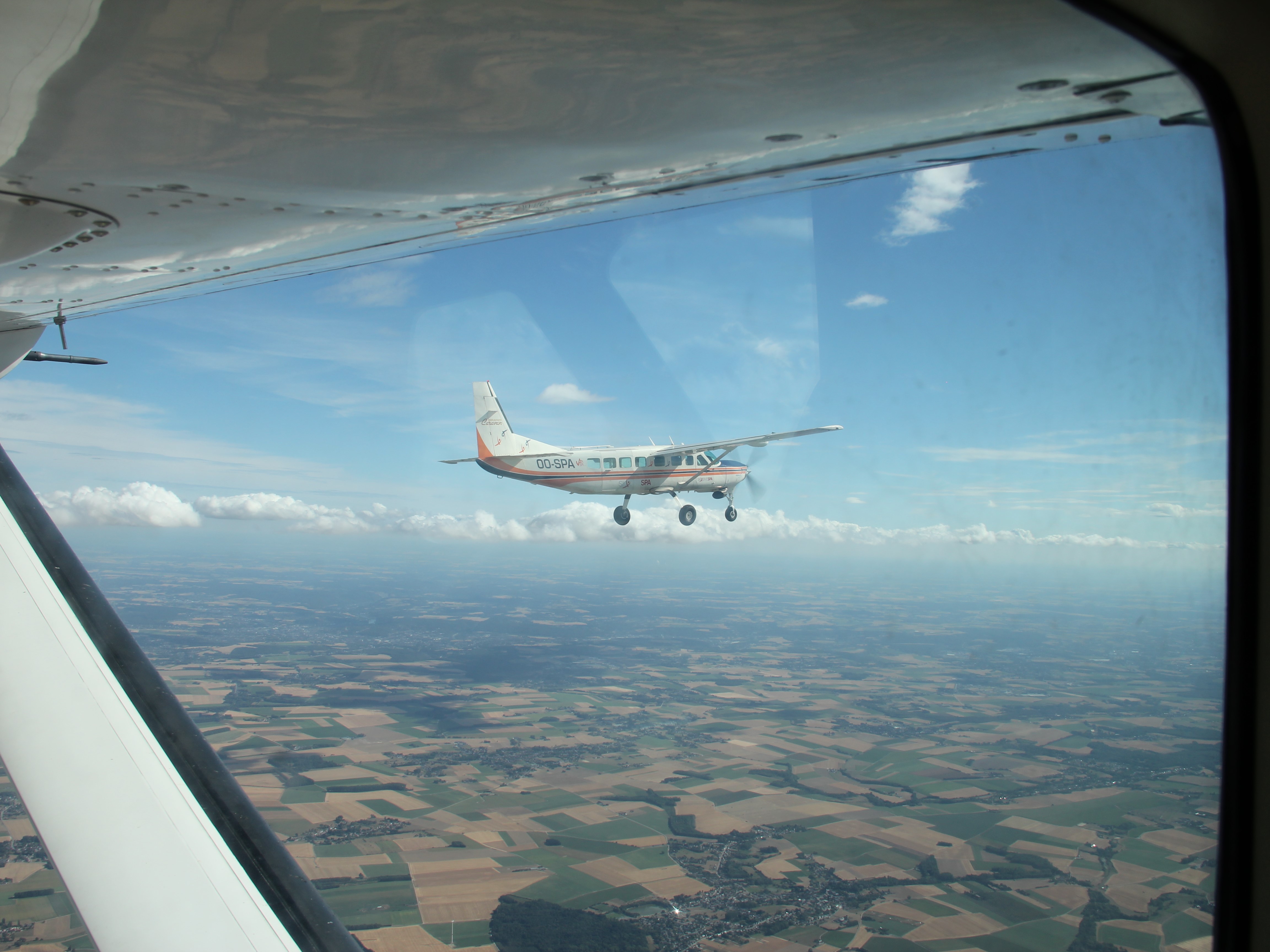, width=\linewidth}}
    \end{minipage}
    \hfill
    \begin{minipage}[b]{.30\linewidth}
        \centering
        \centerline{\epsfig{figure=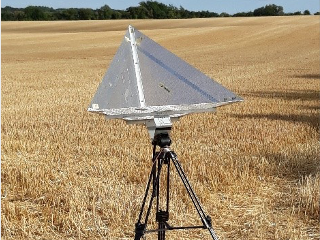, width=\linewidth}}
    \end{minipage}
    \caption{MetaSensing BV's SAR system on-board one of the two CESNA Gran Caravan airplanes (left) and a picture of one airplane taken from the other while flying in the across-track (XTI) configuration (middle), and a corner reflector in the field (right).}%
    \label{fig:planes_and_CR}
\end{figure}

Mono- and bistatic airborne radar data were acquired simultaneously by two left-looking L-band SAR operated by MetaSensing BV on-board two CESNA Gran Caravan airplanes specifically adapted for the mission. The planes and radar system are shown in \cref{fig:planes_and_CR}. The radar operated in frequency modulated continuous wave mode (FMCW) at a central frequency of \SI{1.3}{\giga\hertz}, with a pulse repetition frequency (PRF) at \SI{1004}{Hz}, and a sampling frequency of \SI{50}{\mega\hertz}. The authorized signal bandwidth, allocated by the Belgian Institute for Post and Telecommunications (BIPT), was limited to \SI{50}{\mega\hertz}.
The sensors, capable of providing imaging with spatial resolution up to \SI{1}{\meter}, were equipped with two flat antennas: a squared one and a rectangular one, with a nominal antenna look-angle of \SI{45}{\degree}, and a beamwidth of \SI{40}{\degree} in elevation and respectively \SI{40}{\degree} and \SI{20}{\degree} in azimuth.
They were synchronized through special techniques based on a dedicated high-accuracy GPS-disciplined oscillator (GPSDO) system. Additionally, high-end Global Navigation Satellite System (GNSS)/Inertial Navigation System (INS) devices were installed on the sensors to be able to track their navigation and attitude precisely. Each sensor transmitted and received in a fully-polarimetric ping-pong mode on a chirp-to-chirp basis, allowing both sensors to simultaneously collect mono- and bistatic SAR images depending on the preferred configuration. 

Radar data were collected during a series of five flight missions, labeled F1 to F5, between 31 May (F1) and 10 September 2018 (F5), with a temporal baseline of approximately one month. \Cref{tab:dates} lists the dates of these flights.

\begin{figure}[h]
    \centering
    \includegraphics[width=.7\linewidth]{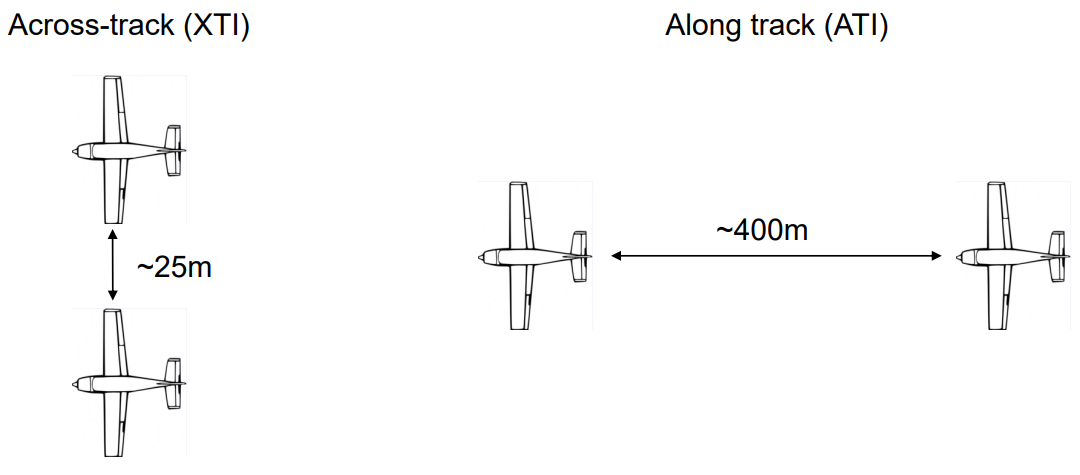}
    \caption{Nadir view of the theoretical XTI (left) and ATI (right) bistatic flight configurations.}
    \label{fig:BELSAR_flight_config}
\end{figure}

\begin{figure}[h]
    \centering
    \includegraphics[width=.65\linewidth]{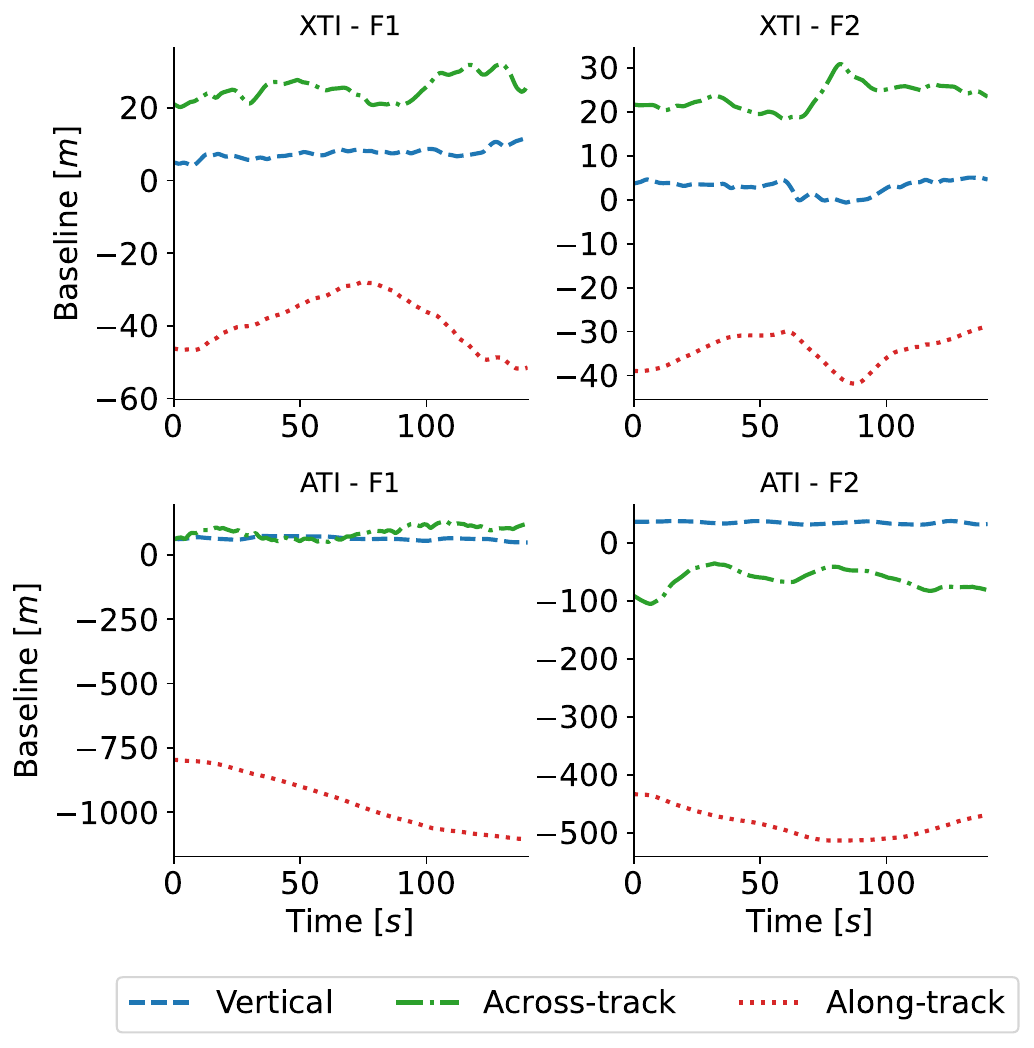}
    \caption{Actual vertical, across-track, and along-track baselines between the two SAR sensors over Zs for the first (F1; left) and second (F2; right) flight missions in the XTI (top) and ATI (bottom) bistatic configurations.}
    \label{fig:BELSAR_baselines}
\end{figure}

On each flight mission, mono- and bistatic data were acquired by flying the airplanes in close formation in two different bistatic geometries, i.e., in across-track (XTI) and in along-track (ATI) configuration.
One aircraft flew steadily at \SI{2500}{\meter} above mean sea level, while the other positioned itself in two distinct locations behind the first.
\Cref{fig:BELSAR_flight_config} depicts the nominal XTI and ATI flight configurations. The actual XTI and ATI baselines were about \SI{35}{\meter} and \SI{450}{\meter}, respectively, for all flight missions except for the first one (F1) when the along-track baseline was about \SI{900}{\meter} and the horizontal separation was in a \SIrange{60}{80}{\meter} range in the ATI configuration. The actual baselines are shown on \cref{fig:BELSAR_baselines} for the first two flight missions, F1 and F2.
These acquisition configurations were established on the basis of ESA's requirement to reproduce the configuration planned for the SAOCOM-CS satellite mission \cite{gebert2014saocom}.

\begin{figure}[h]
    \centering
    \includegraphics[width=.65\linewidth]{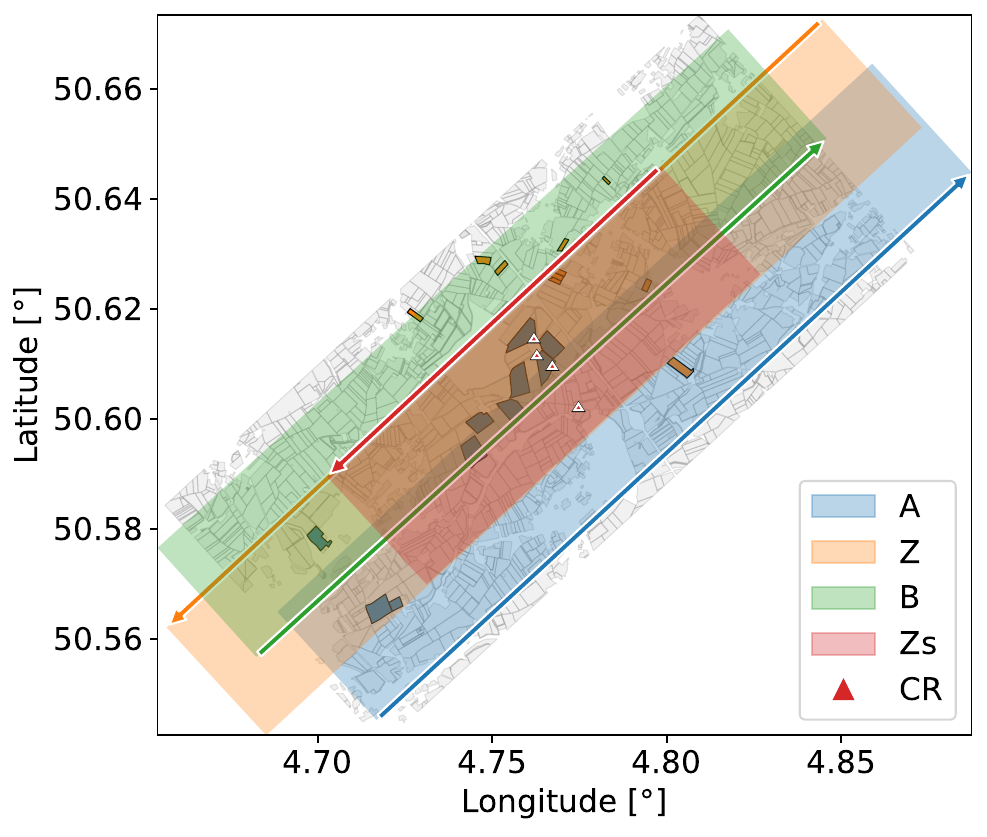}
    \caption{Flight track design in alternating opposite directions and corner reflectors localization within the subset area, Zs. The arrows indicate the trajectory of the left-looking sensors for each track.}
    \label{fig:BELSAR_tracks}
\end{figure}

Three partially overlapping passes were necessary to cover the area of interest. These passes divided the \SI{4.5}{\kilo\meter} wide area depicted in \cref{fig:BELSAR_tracks} into three main tracks labeled Alpha (A), Zulu (Z), and Bravo (B). 
Four trihedral corner reflectors with a side-length of \SI{75}{\centi\meter}, one of which is shown in \cref{fig:planes_and_CR}, were deployed in a fourth, shorter track labeled Zulu short (Zs). They were placed in a row in the across-track direction, and used to produce a geometrical reference as well as the point spread function (PSF) for monostatic acquisitions. Their elevation angle was adjusted for each flight mission, depending on its altitude. Their positions were measured with a precision of \SI{1}{\meter} using a GNSS receiver. 
The complete set of tables describing the tracks and corner reflector installation can be found in the final report of the BELSAR-Campaign project \cite{belsar2018}.

Radar data were processed using the Global Back-Projection (GBP) algorithm, which also handles motion compensation. They were delivered as $\sigma$-calibrated single look complex (SLC) focused SAR data in ground range geometry, with a ground resolution of \SI{1}{\meter} in azimuth and \SI{4}{\meter} in slant range, after Hann windowing with a \SI{50}{\mega\hertz} transmitted pulse.
The SAR images were geo-referenced and co-registered on a sub-pixel level based on the absolute position accuracy of the navigation data, i.e., \SI{0.75}{\meter}.
The implemented SAR processing chain is shown in \cref{fig:SAR_processing_chain}. 

Geometric calibration was performed using the corner reflectors. 
For bistatic data, two range delays had to be defined, one for each system. One of the range delays was adjusted relatively to the absolute calibrated one according to a global offset obtained from a coherence-based fine sub-pixel co-registration. 

The same radiometric calibration was applied to both mono- and bistatic data. The absolute calibration constant, $K$, was evaluated for each flight mission using the four corner reflector responses. Antenna pointing direction and incidence angles were computed using post-processed navigation data and an external digital elevation model (DEM). 
In the XTI configuration, the corner reflectors were assumed to behave in the same way for both mono- and bistatic sensors.
In the ATI configuration, corner reflectors were not visible on bistatic images. The radiometric offset for bistatic ATI data was therefore based on monostatic ATI and bistatic XTI data. 

Finally, a polarimetric calibration was also applied following the procedure described in Fore \textit{et al.} \cite{fore2015uavsar}. Data were calibrated for co-pol and cross-pol channel imbalances and phase bias, i.e., amplitude and relative phase differences between co-polarization channels, at both transmission and reception using the corner reflectors. 
Cross-pol imbalance and phase bias were estimated using the ratio of the averaged cross-pol responses from a large number of pixels.

\subsection*{In situ measurements}

Vegetation and soil bio-geophysical variables were measured in the ten maize and ten winter wheat fields by two teams from the Earth and Life Institute at UCLouvain and the Hydro-Climate Extremes Lab (H-CEL) at Ghent University quasi-simultaneously with the airborne SAR acquisitions. The in situ measurements were performed with a delay of maximum six days with respect to the airborne SAR missions. \Cref{tab:dates} contains the dates of the different measurements.

The sampled fields---labeled with a letter, M for maize and W for wheat, and a number ranging from 1 to 10---are located at the BELAIR Hebania test site, in the Hesbaye region of Belgium. This site belongs to the global Joint Experiment for Crop Assessment and Monitoring (JECAM) network. The location of the site and the fields is shown in \cref{fig:BELSAR_AOI}. The area corresponds to a typical landscape of intensive agriculture in Belgium. The fields are relatively large, flat, and homogeneous, with a uniform topsoil texture of silt loam. The major crops in the area are wheat, potatoes, beets, and maize.

At the time of the first acquisition, winter wheat and maize crops were already in place. Harvesting took place between the second (F2) and third (F3) acquisitions for winter wheat, and between the third (F3) and last (F5) for maize. Some SAR acquisitions were therefore conducted over bare fields. It should be noted that maize stalks were still present on the fields after harvest (see Table \ref{tab:dates}).

Finally, due to the overlap in the flight tracks, certain fields were imaged several times on the same date in both flight configurations, i.e., ATI and XTI. Other fields, however, were imaged only during certain flight missions due to the length of the tracks varying between flight missions, and two maize fields namely M9 and M10, were never imaged by the airborne SAR system.

\subsubsection*{Vegetation}

Sowing density, canopy cover fraction and GAI, plant height, plant development stage, wet biomass and vegetation dry matter content were measured by the team from UCLouvain to characterize the vegetation canopy.
The measurement protocols for maize and winter wheat are similar, but adaptations were applied in consideration of the specificities of both crops.
Measurements were recorded starting in May (F1) until the harvest, which occurred at different dates in the sampled fields, i.e., between June (F2) and end of July (F3) in winter wheat and between end of July (F3) and September (F5) in maize. 

In each field, three homogeneous and representative plots were determined before the campaign based on high-resolution Pleiades images and normalized difference vegetation index (NDVI) profiles derived from them. The plots were marked with flags and geolocalized at the beginning of the growing season. They were named after the field they belong to with the letter a, b, or c attached, e.g., M2a for the first plot of the maize field M2. Field measurements of vegetation were made in the center of these plots around each airborne SAR acquisition. They were chosen to be at least \SI{30}{\meter} from the edges of the field and from each other. The plots were squares with a side length of \SI{15}{\meter} in maize fields and for winter wheat the side length was \SI{25}{\meter}, i.e., the distance between two tractor lines. Correct use of the values provided in the vegetation dataset involves averaging the values of the variables in the three plots to obtain a value representative of the field.

Plant density was measured once at the beginning of the growing season by measuring both interline and interplant distances. To obtain the interline distance in winter wheat fields, the distance between six rows was measured three times per plot for different rows. It was then divided by five, the number of intervals between six rows and averaged. In maize fields, row spacing was measured directly between two rows three times per plot across different rows. The interplant distance was derived in the same way in both crops by counting the number of plants in a one-meter length of sowing row three times per plot.

The canopy cover fraction (FCover), i.e., the fraction of ground surface covered by green plant material, and the GAI, i.e., half of the total area of green plant in the canopy per unit of horizontal ground surface area, were measured by means of digital hemispherical photography (DHP) processed with the CAN-EYE software \cite{weiss2017can_eye}. 
Note that the GAI is the biophysical variable actually estimated in the remote sensing of the leaf area index (LAI) \cite{duveiller2011retrieving}.
In each plot, ten photos were taken with a nadir view approximately \SI{1}{\meter} above the canopy using a system consisting of a camera equipped with a hemispherical lens mounted on a \SI{3}{\meter} telescopic L-shaped pole. Between each photo, the operator walked five steps, or approximately \SI{3.75}{\meter}. The photos were taken each time before the operators entered the plot to avoid altering the measurement. 

The phenological development of the crops was reported on the BBCH scale \cite{meier1997growth}. Only one value was recorded per plot--in case plants in the same plot were at different stages of development, the BBCH stage of the majority of the plants was selected. 

The height of the plants was measured with a ruler from the ground level to its highest part, i.e., leaf, flower, ear or panicle depending on the crop and the stage of development, without extending it manually. For each run, nine plants were measured in each plot to derive a field average. 

\begin{figure}[!t]
    \begin{minipage}[b]{.30\linewidth}
        \centering
        \centerline{\epsfig{figure=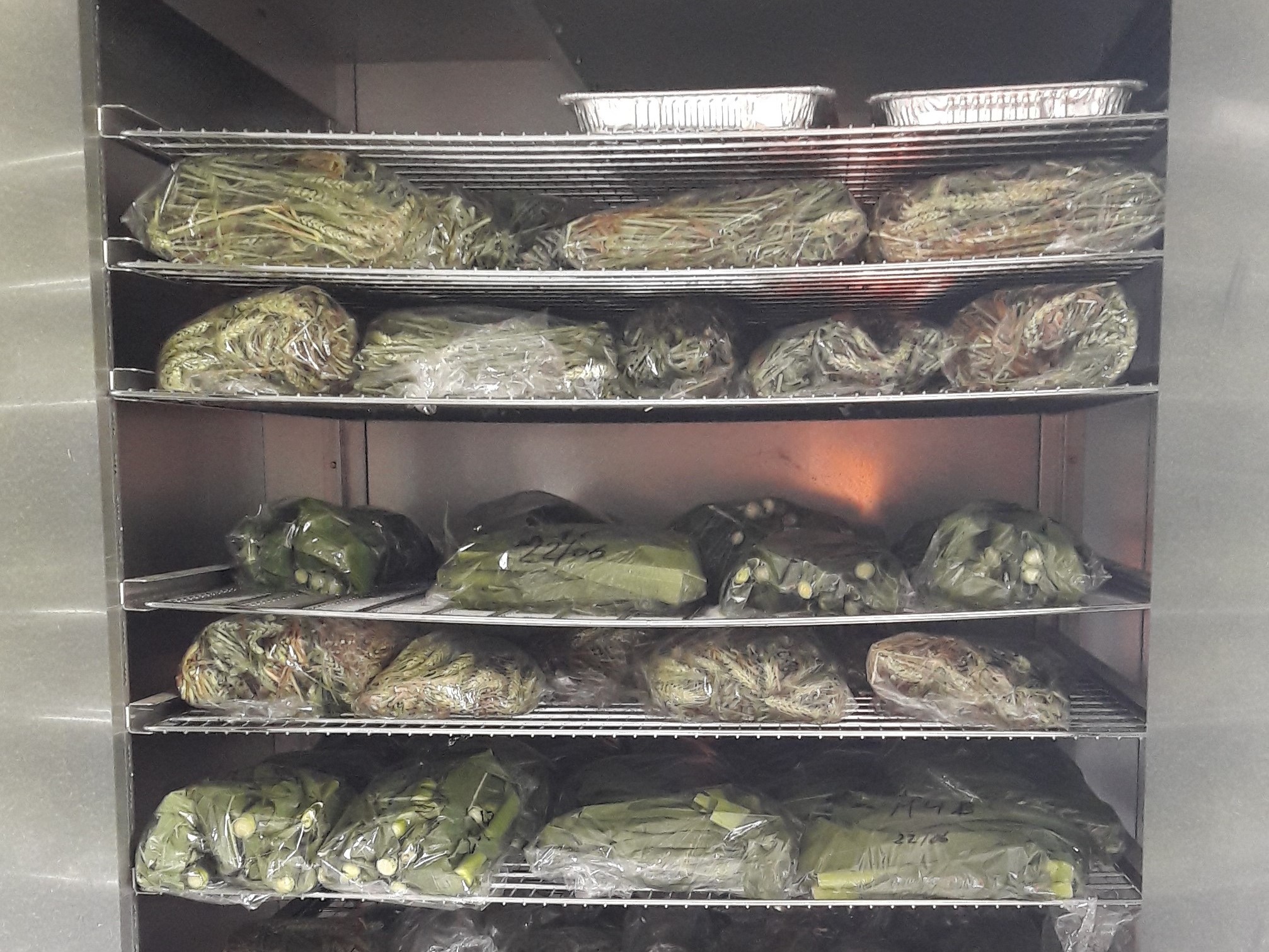, width=\linewidth}}
    \end{minipage}
    \hfill
    \begin{minipage}[b]{.30\linewidth}
        \centering
        \centerline{\epsfig{figure=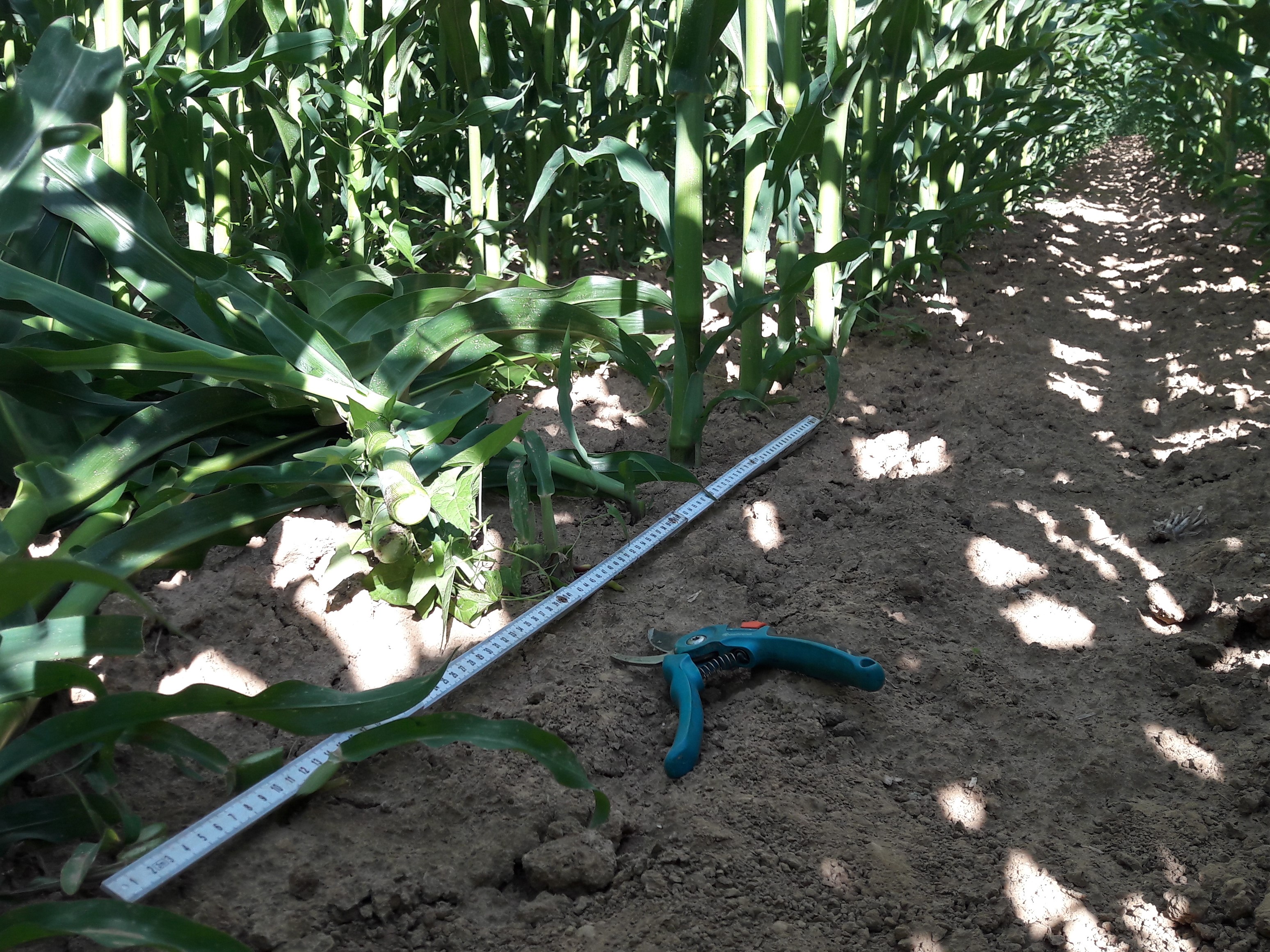, width=\linewidth}}
    \end{minipage}
    \hfill
    \begin{minipage}[b]{.30\linewidth}
        \centering
        \centerline{\epsfig{figure=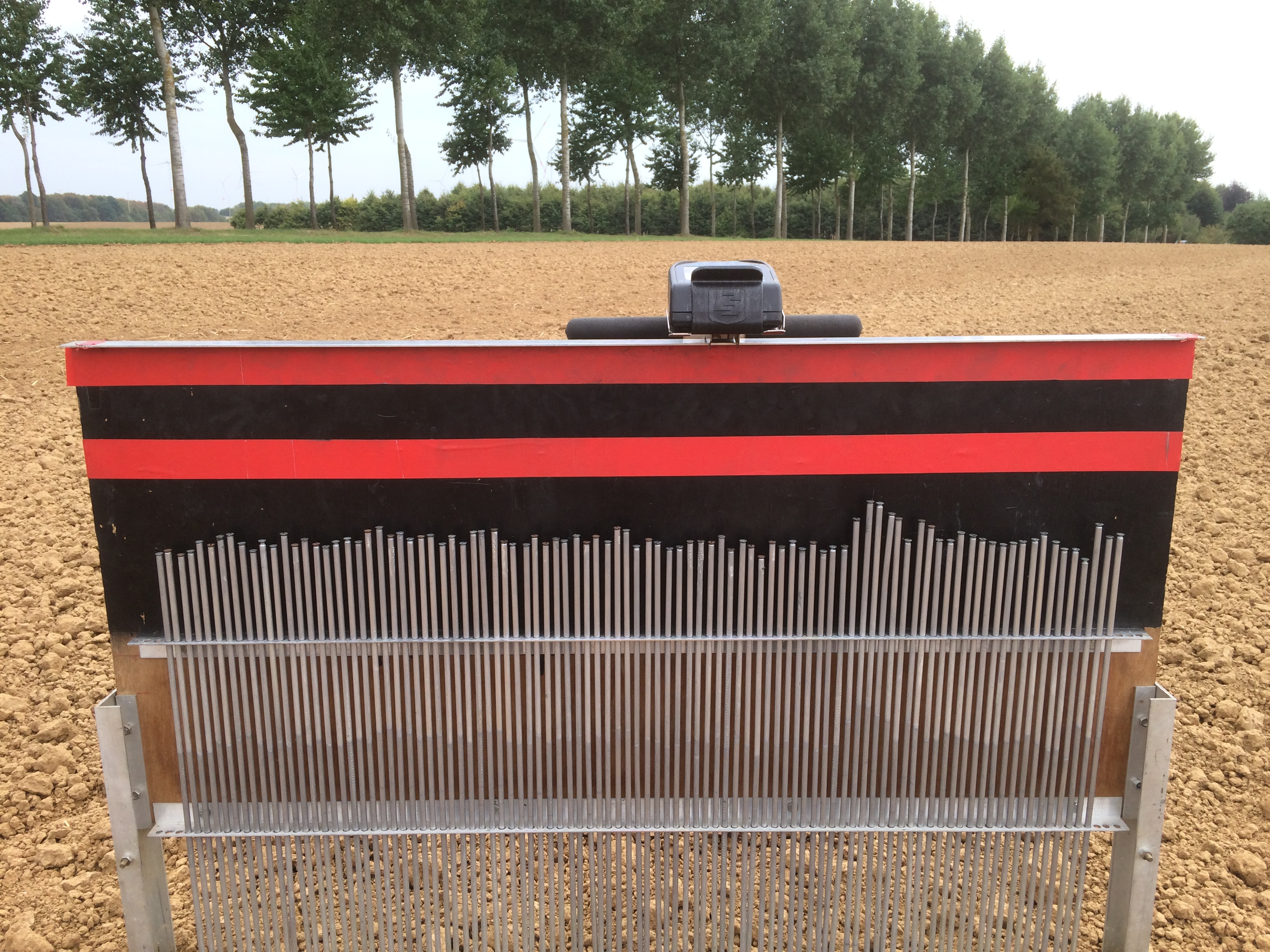, width=\linewidth}}
    \end{minipage}
    \caption{Oven drying of the plant subsamples (left), crop cutting along a \SI{1}{m} row (middle) and pin profilometer (right).}%
    \label{fig:field_measurements}
\end{figure}

The wet biomass and the dry matter content of the crop vegetation were measured by destructive sampling. In each plot, all the plants along three \SI{1}{\meter} rows were cut and then weighed in the field to determine their total fresh weight. The three harvested rows were randomly selected diagonally from each other. A subsample of the cut plants was then randomly selected, weighed and stored in a micro-perforated plastic bag which was then transported to UCLouvain for oven drying at \SI{60}{\celsius} for \SI{72}{\hour}. The dry weight of the subsample was then measured to obtain the dry matter content of the plants. There were no weeds in the plots. Pictures of the subsamples in the oven and of the crop cutting are shown in \cref{fig:field_measurements}.

\subsubsection*{Soil}

The team from Ghent University monitored three soil variables: bulk density, surface soil moisture and surface roughness. Samples were recorded over the entire surface of the studied fields.

Bulk density was measured in all fields alongside the first airborne SAR acquisition (F1) using Kopecky rings. Five samples were taken per field, in order to compute field average values and within-field standard deviations. Additional samples were taken if the bulk density changed due to tillage operations in later flight missions. 

Volumetric soil moisture samples were taken in all ten winter wheat and ten maize fields concurrently with each flight airborne SAR acquisition. During the first acquisition, in May (F1), no samples could be taken in fields W8, W9 and W10, because tillage operations took place during the acquisition time. Soil moisture was monitored using Time Domain Reflectometry (TDR) sensors with \SI{11}{\centi\meter} rods. At least ten locations per field were monitored, with three repetitions per location. Field average volumetric soil moisture was then calculated, by averaging all measurements within each reference plot. 

A pin profilometer, shown in \cref{fig:field_measurements}, of \SI{1}{\meter} length with a spacing of \SI{1}{\centi\meter}, was used to measure soil surface roughness. Roughness samples were taken in all ten winter wheat fields concurrently with every flight campaign, except for field W7 on F4, in August, when tillage operations were taking place at the time of sampling. Surface roughness samples in maize fields could only be taken on F1 and F2. On F3, the maize vegetation was too dense to take samples, resulting in missing values. Between July (F3) and August (F4) campaigns, M1, M6 and M8 were harvested, which made it possible to acquire roughness measurements on F4 and F5. Between F4 and F5, three additional maize plots were harvested (M3, M4 and M9). This way, six maize plots could be sampled on F5. 
The roughness measures were taken in two directions, i.e., along and across the direction of tillage. Pictures of the pin profilometer in both directions were taken and then digitized to correct for tilted pictures and to allow determination of correlation lengths ($l$) and root-mean-square (RMS) heights ($s$). Both were calculated according to the procedure described by Davidson \textit{et. al}\cite{davidson2003joint}. $s$ was calculated as

\begin{equation}
    s=\sqrt{\frac{\sum_{i=1}^{N}z^{2}_{i}}{N-1}},
\end{equation}

\noindent with $N$ the number of profile points and $z_{i}$ the surface height of the profile points.
To estimate $l$, the normalized autocorrelation function $A(\tau)$ was first defined as

\begin{equation}
    A(\tau)=\frac{\sum_{i=1}^{N-k}z_{i}z_{i+k}}{\sum_{i=1}^{N}z_{i}^{2}},
\end{equation}

\noindent then $l$ was obtained by linearly interpolating between two lags, $\tau_1$ and $\tau_2$, that bracket $A(\tau) = e^{-1}$ as

\begin{equation}
    l=\tau_{1}+(e^{-1}-A(\tau_{1}))\frac{\tau_{2}-\tau_{1}}{A(\tau_{2})-A(\tau_{1})}.
\end{equation}

\noindent During each acquisition, at least five profiles per field and along each direction were taken, and for each field that was sampled, the average and standard deviation of the correlation length and root-mean-square height were determined. 


\section*{Data Records}



The BELSAR dataset is available upon request at \hyperlink{https://doi.org/10.5270/ESA-bccf2d9}{https://doi.org/10.5270/ESA-bccf2d9}. 
In it, the SAR, vegetation and soil data are supplied in several files that have been combined into an integrated dataset available at \hyperlink{https://figshare.com/s/e6e480924a1021a42028}{https://figshare.com/s/e6e480924a1021a42028}.

\subsection*{SAR data}

The airborne SAR images can be found in the \textit{RadarData} folder. 
The SAR data are stored in \num{320} NetCDF files---\num{2} sensors (SAR, i.e., monostatic, and BISAR, i.e., bistatic) $\times$ \num{5} flights missions (F1, F2, F3, F4, and F5) $\times$ \num{4} flight tracks (A, Z, B, and Zs) $\times$ \num{2} bistatic configurations (ATI and XTI) $\times$ \num{4} polarizations (HH, HV, VH, and VV)---and delivered with all ancillary metadata that are necessary for further processing and analysis, including antenna pattern, navigation data, digital elevation model, and position of the focused pixels in geographical coordinates (WGS84) \cite{belsarNetCDF2018}. 

The flight track corresponding to each SAR acquisition is provided in \textit{BELSAR\_airborne\_acquisitions\_map\_track\_datetime.csv}.

\subsection*{In situ data}

The vegetation and soil datasets can be found in the \textit{Insitu} folder, in which the main files are:
\begin{itemize}
    \item \textit{BELSAR\-\_agriculture\-\_database.xlsx}
    \begin{itemize}
        \item \textit{BelSAR\_maize} tab -- Vegetation biophysical variables recorded in each plot of each maize field
        \item \textit{BelSAR\_wheat} tab -- Vegetation biophysical variables recorded in each plot of each winter wheat field
    \end{itemize}
    \item \textit{BELSAR\-\_soil\-\_bulkdensity\-\_database.xlsx}
    \begin{itemize}
        \item \textit{Field\_average\_bulkdensity} tab -- Mean bulk density for each field [\si{\gram\per\cubic\centi\meter}]
        \item \textit{Field\_std\_dev\_bulkdensity} tab -- Standard deviation of the bulk density for each field [\si{\gram\per\cubic\centi\meter}]
    \end{itemize}
    \item \textit{BELSAR\-\_soil\-\_moisture\-\_database.xlsx}
    \begin{itemize}
        \item \textit{Field\_average} tab -- Mean volumetric soil moisture for each field [\si{\percent}]
        \item \textit{Field\_std\_dev} tab -- Standard deviation of the volumetric soil moisture for each field [\si{\percent}]
        \item \textit{Raw\_data} tab -- Overview of the raw data: latitude, longitude (WGS84), volumetric soil moisture [\si{\percent}], period [\si{\micro\second}], attenuation and permittivity of all TDR samples. For F3, F4 and F5, the latter three can be missing due to sensor failure. 
    \end{itemize}
    \item \textit{BELSAR\-\_soil\-\_roughness\-\_database.xlsx}
    \begin{itemize}
        \item \textit{Field\_average\_corr\_length} tab -- Mean correlation length for each field [\si{\centi\meter}]
        \item \textit{Field\_average\_RMSheight} tab -- Mean root-mean-square height for each field [\si{\centi\meter}]
        \item \textit{Field\_std\_dev\_corr\_length} tab -- Standard deviation of the correlation length for each field [\si{\centi\meter}]
        \item \textit{Field\_std\_dev\_RMSheight} tab -- Standard deviation of the root-mean-square height for each field [\si{\centi\meter}]
    \end{itemize}
\end{itemize}

\subsection*{Integrated dataset}

To facilitate their exploitation, the SAR and in situ datasets have been integrated into a single dataset, \textit{BELSAR\_fields\_integrated\_db.csv}.
The integrated dataset includes zonal statistics of the radar measurements as well as the bio- and geophysical variables of soil and vegetation for each maize and winter wheat field.
Each entry in the table corresponds to a given field and image in the SAR dataset. 
The codes used to generate the integrated dataset are provided in the same folder.

\section*{Technical Validation}


\subsection*{SAR data calibration}

\begin{figure}[h]
    \centering
    \includegraphics[width=.65\linewidth]{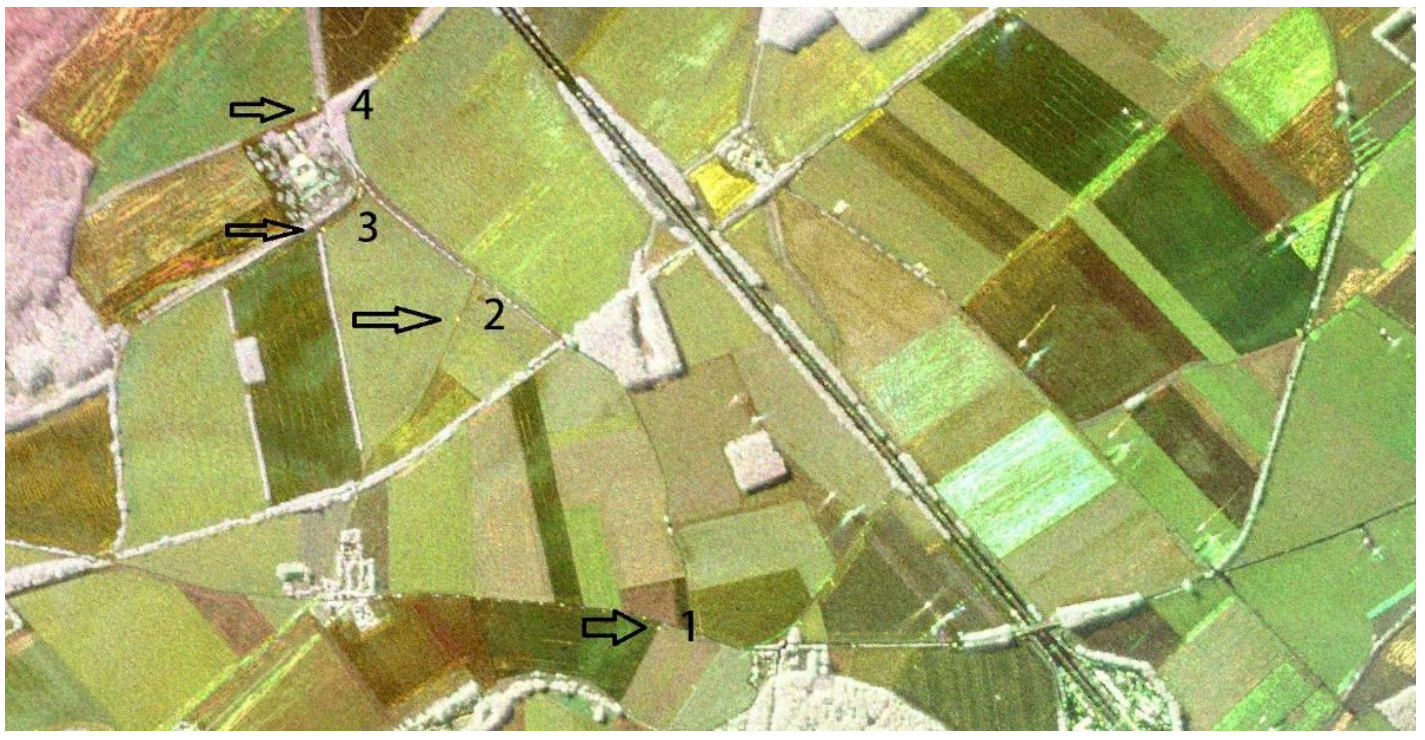}
    \caption{Polarimetric RGB composite image around the corner reflectors (indicated by the arrows), where red, green and blue correspond to HH, VV, and HV, respectively.}
    \label{fig:BELSAR_pol_image}
\end{figure}

The SAR data were geometrically, radiometrically, and polarimetrically calibrated based on the response of the four corner reflectors deployed in track Zs. 
\Cref{fig:BELSAR_pol_image} shows the polarimetric RGB composite image in the vicinity of the corner reflectors. 
Their responses show that the resolution of the images are within \SI{1}{\meter} resolution in azimuth and \SI{4}{m} in slant range, as intended, and that the corners are correctly geolocated, with an accuracy of \SI{0.75}{\meter}.
Their polarimetric signatures and impulse response function also indicated a good isolation between the polarization channels at antenna level. Cross-talk was therefore considered negligible, as also attested by their appearance as yellow spots on the RGB composite, i.e., only HH and VV response with no significant imbalances.
As for the radiometric calibration of the bistatic data in the ATI flight configuration, however, given that the corner reflectors were not visible on the images, their radiometric offset was determined using monostatic ATI and bistatic XTI data. This may have led to an imbalance between bistatic and monostatic data, or from one flight to the next, in the bistatic ATI data.
A complete quality assessment can be found in the BELSAR-Campaign project final report \cite{belsar2018}.

\subsection*{Interferometric SAR}

The potential of the airborne data for interferometric SAR (InSAR) was affected by very low interferometric coherence. 
Strong temporal coherence losses were observed in double-pass monostatic pairs, i.e., between the different flight missions, together with a significant geometric decorrelation due to large orthogonal baselines as compared to the critical baseline. Coherence was also low for single-pass bistatic interferometric pairs, despite the short perpendicular baselines. The short signal bandwidth, restricted to \SI{50}{\mega\hertz} by the BIPT, has certainly played a significant role on geometric decorrelation. This effect was strengthened by the relatively low flight altitude. 
Furthermore, in addition to temporal and geometric decorrelation, coherence might have been affected, in bistatic configuration, by a lack of synchronization resulting from using two different clocks for the transmitter and receiver systems, compared with a monostatic configuration where the unique transmitter-receiver system operates with absolute time and phase references. The consequences of mis-synchronization errors are positioning and phase errors in the output SLC images, giving rise to additional azimuthal fringes in the interferogram with a detrimental effect on coherence. A solution to this last issue would be to implement a synchronization algorithm that minimizes the impact of residual plane motion and mitigates the fringe mis-synchronization by a phase calibration. A multisquint based correction algorithm is proposed by de Macedo \textit{et al.} \cite{macedo2021} to solve this.

\subsection*{Vegetation and soil measurements}

\subsubsection*{Vegetation}
\begin{figure}[h]
    \centering
    \includegraphics[width=\linewidth]{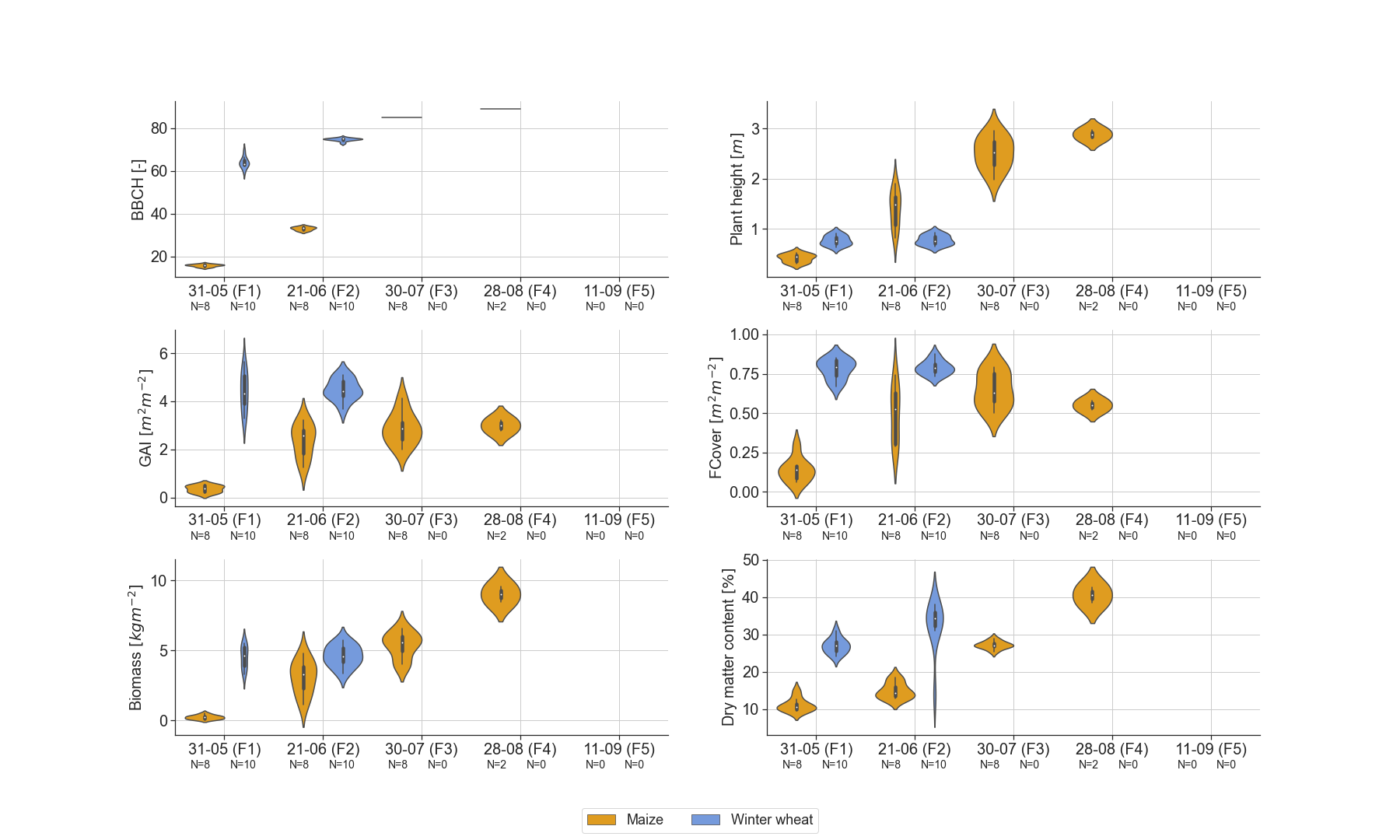}
    \caption{Violin plots of vegetation biophysical variables measured in situ, i.e., phenological development stage (BBCH; top-left), plant height (top-right), green area index (GAI; middle-left), canopy fraction cover (FCover; middle-right), wet biomass (bottom-left), and dry matter content (bottom-right), for maize (orange) and winter wheat (blue) for all flights (F1-F5). The number of fields that were monitored during each flight is denoted by $N$.}
    \label{fig:BELSAR_violin_vegetation}
\end{figure}

The vegetation measurements were performed in accordance with the guidelines laid down by JECAM \cite{hosseini2018soil}. Violin plots of in situ measured vegetation biophysical variables, i.e., phenological development stage (BBCH), plant height, GAI, FCover, wet biomass, and dry matter content, are shown in Figure \ref{fig:BELSAR_violin_vegetation}.

Maize was observed between BBCH stages \num{15} and \num{89}, i.e. from the leaf development stage (five leaves) to the fully ripe stage, and winter wheat between stages \num{59} and \num{76}, i.e. from the end of the heading to medium milk stage. Few data, in a narrow range of values, are available on winter wheat because the campaign started late in an unusually warm and dry growing season, leading to an early maturation and harvest.

The mean plant height observed in the different fields ranges from \SI{0.42}{\meter} (with a standard deviation of \SI{0.08}{\meter}) to \SI{2.89}{\meter} (\SI{0.09}{\meter} standard deviation) in maize. These values are in accordance with maize plant height measurements over a loamy test site in Belgium \cite{blaes2006c}. For winter wheat plant height ranges from \SI{0.77}{\meter} (\SI{0.09}{\meter} standard deviation) to \SI{0.78}{\meter} (\SI{0.09}{\meter} standard deviation).

The mean FCover and GAI over all maize fields range respectively from \SI{0.14}{m^2m^{-2}} (\SI{0.71}{m^2m^{-2}} standard deviation) on F1 to \SI{0.55}{m^2m^{-2}} (\SI{0.30}{m^2m^{-2}} standard deviation) on F4 with a maximum at \SI{0.65}{m^2m^{-2}} (\SI{0.10}{m^2m^{-2}} standard deviation) on F3 and from \SI{0.36}{m^2m^{-2}} (\SI{0.14}{m^2m^{-2}} standard deviation) on F1 to \SI{3.0}{m^2m^{-2}} (\SI{0.24}{m^2m^{-2}} standard deviation) on F4, the latter also being the maximum mean GAI value. In winter wheat, both ranges are very limited, from \SI{0.78}{m^2m^{-2}} (\SI{0.06}{m^2m^{-2}} standard deviation) to \SI{0.79}{m^2m^{-2}} (\SI{0.04}{m^2m^{-2}} standard deviation) for the FCover and from \SI{4.41}{m^2m^{-2}} (\SI{0.75}{m^2m^{-2}} standard deviation) to \SI{4.49}{m^2m^{-2}} (\SI{0.42}{m^2m^{-2}} standard deviation) for the GAI. The small range of values for winter wheat were expected given the late start of the campaign, which implies that winter wheat was already close to maturation on F1. These GAI values for winter wheat at maturation stage, are in line with field-observed GAI over a test site in Wallonia (southern part of Belgium) \cite{de2012estimating}. 

The mean of dry matter content over all fields ranges from \SI{11.18}{\percent} (\SI{1.68}{\percent} standard deviation) to \SI{40.57}{\percent} (\SI{2.18}{\percent} standard deviation) in maize and from \SI{27.01}{\percent} (\SI{1.96}{\percent} standard deviation) to \SI{32.42}{\percent} (\SI{6.51}{\percent} standard deviation) in winter wheat. The maize was not let to dry further because it was intended for silage.

The mean wet biomass over all maize fields ranges from \SI{0.24}{kg/m^2} (\SI{0.14}{kgm^{-2}} standard deviation) on F1 to a maximum mean of \SI{9.01}{kgm^{-2}} (\SI{0.56}{kgm^{-2}} standard deviation) on F4, in line with values found in the study of Blaes \textit{et al.} \cite{blaes2006c} over a maize field in a loamy test site. For winter wheat fields, the mean wheat biomass ranges from \SI{4.53}{kgm^{-2}} (\SI{0.77}{kgm^{-2}} standard deviation) on F1 to a maximum mean of \SI{4.61}{kgm^{-2}} (\SI{0.72}{kgm^{-2}} standard deviation) on F2. 

\subsubsection*{Soil}

\begin{figure}[h]
    \centering
    \includegraphics[width=\linewidth]{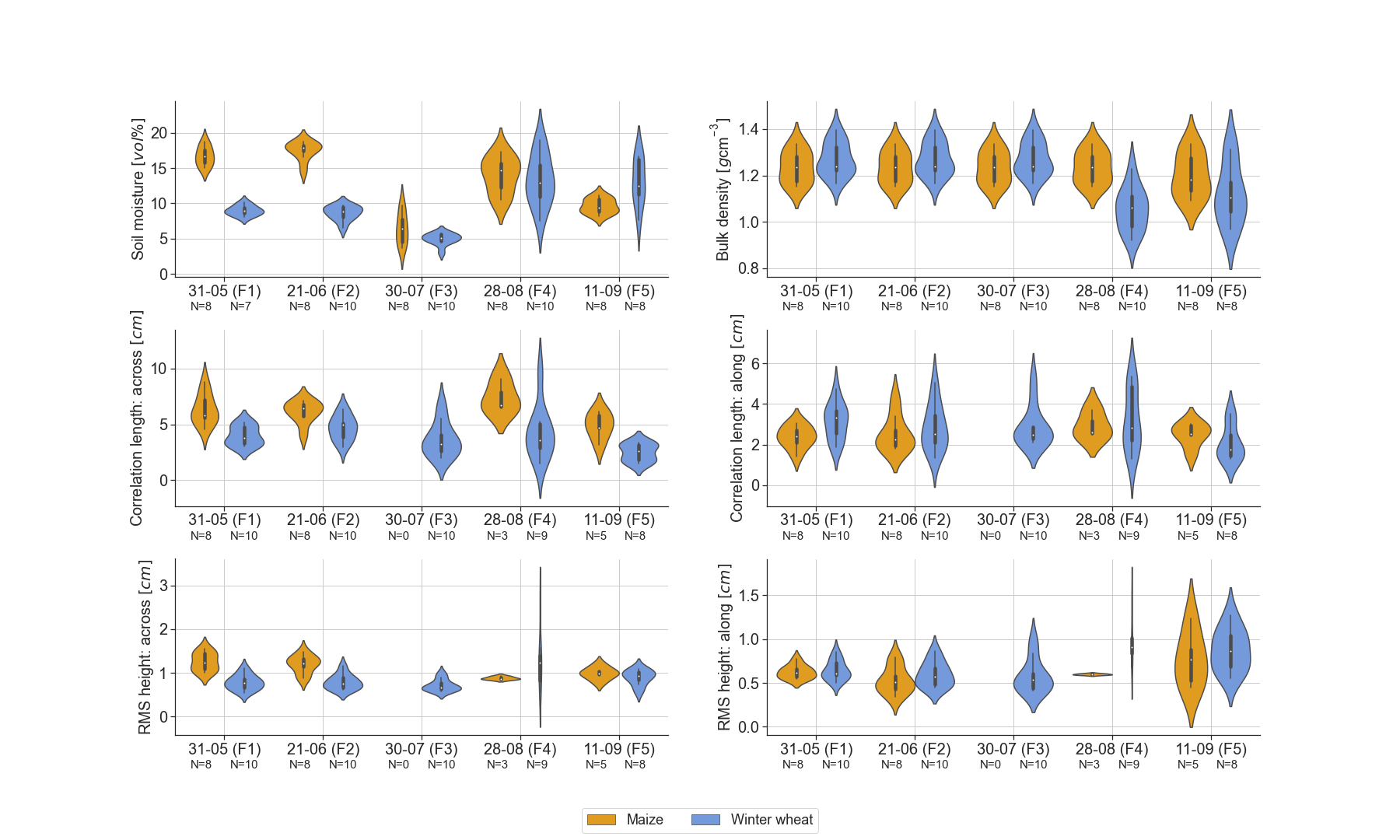}
    \caption{Violin plots of soil geophysical variables measured in situ, i.e., soil moisture (top-left), bulk density (top-right), correlation length across (middle-left) and along (middle-right) tillage direction, root-mean-square (RMS) height across (bottom-left) and along (bottom-right) tillage direction, for maize (orange) and winter wheat (blue) for all flights (F1-F5). The number of fields that were monitored during each flight is denoted by $N$.
    }
    \label{fig:BELSAR_violin_soil}
\end{figure}

Figure \ref{fig:BELSAR_violin_soil} shows violin plots for the soil geophysical variables, i.e., soil moisture, bulk density, correlation length across, correlation length along, root-mean-square height across, and root-mean-square height along, for maize and winter wheat fields.

Bulk density was measured in all winter wheat and maize fields during the first flight campaign (F1), with a mean bulk density of \SI{1.27}{\gram\per\cubic\centi\meter} (with an average within field standard deviation of \SI{0.079}{\gram\per\cubic\centi\meter}) for the winter wheat fields and \SI{1.24}{\gram\per\cubic\centi\meter} (with an average within field standard deviation of \SI{0.065}{\gram\per\cubic\centi\meter}) for the maize fields. Similar bulk density values have been reported in the study of van der Bolt \textit{et al.} \cite{van2020bodemverdichting} where agricultural fields in Flanders (Belgium) were sampled, including sites with a texture of silt loam. Due to tillage operations that took place between F3 and F4 on the winter wheat fields (except W7), additional soil samples were taken during the field campaign coincident with F4 to determine bulk density values after tillage operations. On F5, additional samples were taken in winter wheat fields W2 and W7 and in maize fields M1, M3 and M4. The plots show that bulk density generally decreases after tillage operations. 

The field average volumetric soil moisture values range from \SIrange{3.03}{18.94}{\volpercent} for winter wheat fields and from \SIrange{3.65}{18.76}{\volpercent} for maize fields. These values are within the range of soil moisture values reported in the study of Choker \textit{et al.} \cite{choker2017evaluation}, where in situ soil moisture measurements over numerous agricultural plots in Europe  (mainly France, Belgium, and Italy) were acquired. Note that 2018 was marked by an exceptional dry summer in Belgium, which is depicted in the low soil moisture values, especially on F3. The range of within field standard deviations for winter wheat fields is \SIrange{0.68}{4.60}{\volpercent} and for maize \SIrange{1.12}{5.36}{\volpercent}.

The field average root-mean-square heights (correlation length) was measured along and across the direction of tillage. Roughness along the direction of tillage is comparable between winter wheat and maize fields, with a range of respectively \SIrange{0.41}{1.46}{\centi\meter} (\SIrange{1.28}{5.34}{\centi\meter}) for winter wheat and \SIrange{0.32}{1.23}{\centi\meter} (\SIrange{1.42}{4.30}{\centi\meter}) for maize. Across the direction of tillage, the difference is slightly larger, with a range of \SIrange{0.55}{2.63}{\centi\meter} (\SIrange{1.46}{9.69}{\centi\meter}) for winter wheat and \SIrange{0.80}{1.55}{\centi\meter} (\SIrange{3.16}{9.07}{\centi\meter}) for maize. Especially during crop growth stages, the root-mean-square height across the direction of tillage is substantially higher for maize compared to winter wheat, with an average value of \SI{1.23}{\centi\meter} (\SI{6.25}{\centi\meter}) for maize compared to \SI{0.79}{\centi\meter} (\SI{4.38}{\centi\meter}) for winter wheat.

Verhoest \textit{et al.} \cite{verhoest2008soil} summarized possible sources of errors that affect roughness measurements, of which the limited length of the pin profilometer and resolution in both vertical and horizontal directions are the main disadvantages, especially for the estimation of $l$. Roughness measures found in this study are comparable to the ones estimated by Davidson \textit{et al.} \cite{davidson2003joint} for agricultural sites over Europe. A mean $s$ of 0.6 and 1.6 cm (with a standard deviation of 0.3 and 0.7 cm) was estimated for seedbed and harrowed field conditions respectively, which is in line with the BELSAR study. In terms of correlation length, mean values of 3.7 and 3.8 cm (with a standard deviation of 2.6 and 2.9 cm) for respectively seedbed and harrowed field conditions were found \cite{davidson2003joint}, which is in the range of values found here for both maize and winter wheat fields after harvest. The correlation length for maize fields before harvest was substantially larger, which can be explained by the rough seedbed pattern for maize.

\section*{Usage Notes}



The integrated dataset is directly accessible on figshare, while the BELSAR-Campaign data are available online via FTP upon submission of a data access request to ESA's Earth Online service.

With regard to the exploitation of these data, it is advisable to use the integrated dataset instead of the original data as it contains in one table both the zonal statistics of the SAR data and the corresponding vegetation and soil variables for the maize and winter wheat fields. As such, it provides an analysis-ready dataset, thereby greatly facilitating the handling of the BELSAR data for, among others, the development of agricultural or hydrological applications as well as for easy comparison with other comparable datasets.
Note that for certain purposes, it is recommended to apply a negative buffer to the polygons to avoid edge-of-field effects on the radar signal. In this case, the polygons delineating the fields would have to be redrawn and the integrated dataset rebuilt.

\section*{Code availability}


The codes used to produce the integrated dataset from SAR, vegetation and soil data have been uploaded to figshare along with it. These contain a number of tools that can be easily adapted and reused to use the BELSAR data for other purposes.
To rebuild the integrated dataset from the BELSAR-Campaign data, the contents of the ESA repository must first be accessed. Then, once downloaded, running \textit{extract\_mini\_rasters.py} will extract zonal statistics from the SAR data and \textit{integrated\_dataset.py} will match these extracted data to the corresponding in situ vegetation and soil measurements and generate the integrated dataset. For test purposes, these two scripts can be run for one or a series of images by adding their indices, from 0 to 319, as arguments to the python command, e.g., \texttt{python extract\_mini\_rasters.py 1} for the second image and \texttt{python extract\_mini\_rasters.py 0 2} for the first three images.

\bibliography{bibliography}


\section*{Acknowledgements}

This research was conducted in the framework of the BELSAR-Publication project
funded by the STEREO III program of the Belgian Federal Science Policy Office (BELSPO) under contract SR/00/409.

The authors would like to thank Quentin Vandersteen and Jeroen Claessen for the collection of the in situ data, 
Leila Guerriero, Nazzareno Pierdicca, Hugh Griffiths, and Malcolm Davidson for their participation to the BELSAR advisory committee,
and the Service public de Wallonie for providing access to the Land Parcel Information System of Wallonia.

\section*{Author contributions statement}

J.B., E.T., A.O.\ and K.A.C.M.\ collaborated in curating, analyzing and validating the data, as well as writing the manuscript,
J.B.\ and E.T.\ wrote the code, 
J.B., E.T.\ and K.A.C.M.\ plotted the figures,
N.E.C.V.\ and P.D.\ provided guidance and supervision. 
All authors reviewed the manuscript.

\section*{Competing interests}

The authors declare no competing interests.

\section*{Figures \& Tables}

\begin{figure}[h]
    \centering
    \includestandalone[width=.65\linewidth]{Images/flowchart}
    \caption{SAR processing chain}
    \label{fig:SAR_processing_chain}
\end{figure}

\begin{table}[h]
    \centering
    \begin{tabular}{|l|l|l|l|l|}
        \hline
        \textbf{Flight mission} & \textbf{Flight date} & \textbf{In situ measurement date} & \textbf{Maize stage} & \textbf{Winter wheat stage}\\
        \hline
        F1 & 30/05 & 30/05, 31/05, and 05/06 & Leaf development & End of heading \\
        \hline
        F2 & 20/06 & 21/06, 22/06, and 26/06 & Stem elongation & Medium milk \\
        \hline
        F3 & 30/07 & 01/08 and 02/08 & Dough stage & Bare soil \\
        \hline
        F4 & 28/08 & 29/08 and 30/08 & Fully ripe (only M2 and M5) & Bare soil \\
        & & & and bare soil with stalks & \\
        \hline
        F5 & 10/09 & 10/09 and 11/09 & Bare soil with stalks & Bare soil \\
        \hline
    \end{tabular}
    \caption{Dates (in day-month-2018 format) of airborne SAR acquisitions and in situ measurements of vegetation and soil variables, with the corresponding development stages of both crops.}
    \label{tab:dates}
\end{table}

\end{document}

%% file: Images/flowchart_tikz_headers.tex
\usepackage{tikz}
\usetikzlibrary{automata, positioning, arrows, patterns} 
\usetikzlibrary{shapes.geometric, arrows, positioning}

\tikzstyle{oval} = [
draw,
ellipse,
fill=gray!10,
minimum width=60pt,
minimum height=40pt,
text width=50pt
]
\tikzstyle{rect} = [
draw,
rectangle,
fill=gray!10,
minimum width=60pt,
minimum height=40pt,
text width=60pt
]

%% file: Images/flowchart.tex
\begin{tikzpicture}[thick,->,>=stealth,node distance=2cm and 3cm,on grid,text centered]
    \node[oval] (raw_data) {Raw data};
    \node[oval,right=of raw_data] (nav_data) {Navigation data};
    \node[oval,right=of nav_data] (DEM) {DEM};
    \node[rect,below=of raw_data] (ingest) {Ingestion and unpacking};
    \node[rect,below=of nav_data] (nav_post) {Navigation post-processing};
    \node[rect,below=of ingest] (range) {Range compression};
    \node[rect,below=of range] (GBP) {Global Back Projection};
    \node [below=of GBP] (2) {};
    \node[text width=80pt,left=of 2] (bist) {Bi- and monostatic SLC data (ground range geometry)};
    \node[rect,below=of 2] (range_del) {Range delay calibration};
    \node[rect,below=of range_del] (ant) {Antenna pattern and polarimetric calibration};
    \node [right=of ant] (1) {};
    \node[draw,rectangle,fill=gray!10,minimum width=140pt,minimum height=30pt,text width=140pt,below=of 1] (data) {Data and metadata packaging};
    
    \draw (raw_data) edge node [anchor=east] {All 8 channels} (ingest);
    \draw (ingest) edge (range);
    \draw (range) edge (GBP);
    \draw (nav_data) edge (nav_post);
    \draw (nav_post) |- (GBP.10);
    \draw (GBP.255) |- (bist);
    \draw (GBP.255) -- (range_del.105);
    \draw (range_del.75) -- (GBP.285);
    \draw (range_del) edge (ant);
    \draw (nav_post) |- (range_del.10);
    \draw (nav_post) |- (ant.10);
    \draw (nav_post) edge (data);
    \draw (DEM) |- (GBP.350);
    \draw (DEM) |- (range_del.350);
    \draw (DEM) |- (ant.350);
    \draw (DEM) |- (data);
    \draw (ant) |- (data);
\end{tikzpicture}